\documentclass[aps,prc,showpacs,twocolumn,epsfig]{revtex4}
\usepackage[dvips]{graphicx}
\begin{document}

\draft
\title{Formula for proton-nucleus reaction cross section at intermediate
energies and its application}

\author{Kei Iida,$^1$ Akihisa Kohama,$^2$  and Kazuhiro Oyamatsu$^{2,3}$}
\affiliation{
$^1$RIKEN BNL Research Center, Physics Department, Brookhaven National 
Laboratory, Upton, New York 11973-5000, USA\\
$^1$RIKEN (The Institute of Physical and Chemical Research),
2-1 Hirosawa, Wako-shi, Saitama 351-0198, Japan\\
$^3$Department of Media Theories and Production, Aichi Shukutoku
University, Nagakute, Nagakute-cho, Aichi-gun, Aichi 480-1197, Japan}

\date{\today}

\begin{abstract}

     We construct a formula for proton-nucleus total reaction cross 
section as a function of the mass and neutron excess of the target nucleus and 
the proton incident energy.  We deduce the dependence of the cross section on 
the mass number and the proton incident energy from a simple argument 
involving the proton optical depth within the framework of a black sphere 
approximation of nuclei, while we describe the neutron excess dependence by 
introducing the density derivative of the symmetry energy, $L$, on the basis 
of a radius formula constructed from macroscopic nuclear models.  We find that 
the cross section formula can reproduce the energy dependence of the cross 
section measured for stable nuclei without introducing any adjustable 
energy dependent parameter.  We finally discuss whether or not the reaction 
cross section is affected by an extremely low density tail of the neutron 
distribution for halo nuclei.

\end{abstract}

\maketitle

\section{Introduction}
\label{intro}

      The nuclear size and mass are fundamental quantities characterizing
the bulk properties of nuclei.  In fact, the saturation of the binding energy
and density deduced from systematic data for masses and charge radii of
stable nuclei reflects the behavior of the equation of state of nearly
symmetric nuclear matter near the saturation density $n_0$.  Extension of such
data to a more neutron-rich regime opens up an opportunity of probing
the equation of state of asymmetric nuclear matter.  One of the poorly known
parameters characterizing such an equation of state is the density symmetry 
coefficient, $L\equiv 3n_0 dS/dn|_{n_0}$, with the nucleon density $n$ and the 
density dependent symmetry energy $S(n)$.  The sensitivity of $L$ to
the structure and evolution of neutron stars and to nuclear structure and 
reactions has been discussed in Ref.\ \cite{SPLE}.  This parameter is 
expected to be deduced from future systematic data for unstable nuclei because
macroscopic nuclear models yield a clear correlation between the 
root-mean-square (rms) matter radii of very neutron-rich nuclei and $L$
mainly through the $L$ dependence of the saturation density of asymmetric 
nuclear matter \cite{OI}.

      For stable nuclei, electron and proton elastic scattering data have
been obtained systematically, from which the charge and matter density 
distributions can be fairly well deduced 
\cite{Alk:PR,Cha:AP,Igo:RMP,Bat:ANP,Fro:MT}, while for unstable nuclei, not 
only do electron elastic scattering experiments have yet to be performed, but 
also proton scattering data obtained from radioactive ion beams incident on a 
proton target \cite{Alexei,Neumaier1} are very limited.  Instead of relying on
such elastic scattering data, deduction of the rms matter radius of unstable 
nuclei has been often performed through measurements of 
reaction/interaction cross section (e.g., Ref.\ \cite{Ozawa}).  However,
the relation between the reaction/interaction cross section and the 
matter radius is not obvious theoretically and in fact is quite model 
dependent.

     Recently we systematically analyzed the proton elastic scattering and 
reaction cross section data for stable nuclei at proton incident energy of 
$\sim$800--1000 MeV on the basis of a black sphere picture of nuclei 
\cite{BS1,BS2}.  This picture is originally expected to give a decent 
description of the reaction cross section for any kind of incident particle 
that tends to be attenuated in the nuclear interiors.  In fact we showed that 
for proton beams incident on stable nuclei, the cross section of the black 
sphere of radius $a$, which was determined by fitting the angle of the 
first elastic diffraction peak calculated for the proton diffraction by a 
circular black disk of radius $a$ to the measured value, is consistent with 
the measured reaction cross section \cite{BS2}.  We also showed that for 
stable nuclei of mass number $A>50$, the quantity $\sqrt{3/5}a$ agrees with 
the rms matter radius deduced in previous elaborate analyses from the 
proton elastic scattering data within error bars \cite{BS1}.  These salient 
features of the black sphere picture could help deduce the density dependence 
of the symmetry energy from future systematic data on the reaction cross 
section and proton elastic scattering for unstable nuclei.

     In this work we construct a formula for the proton-nucleus total
reaction cross section as a function of the target mass number $A$, the 
proton incident energy $T_p$, and the target neutron excess 
$\delta=1-2Z/A$.  The dependence of the cross section on $A$ and $T_p$, which 
is deduced by combining the black sphere picture of nuclei with a simple
argument involving the proton optical depth, is remarkably consistent with 
the empirical data for stable nuclei at $T_p=100$--1000 MeV.  Bearing in
mind future possible data for unstable nuclei at intermediate energies, we 
then describe the neutron excess dependence by using a radius formula 
\cite{OI} constructed from macroscopic nuclear models in terms of $L$.  As a 
possible application of the formula, we discuss the effect of nuclear 
deformation and neutron halos on the reaction cross section.  We finally
extend the formula to the case of nucleus-nucleus reaction cross section and
compare the result with the existing data.

\section{Formula for proton-nucleus reaction cross section}
\label{fpA}

     We start with a formula for the reaction cross section for protons
on stable nuclei as a function of $A$ and $T_p$.  For this purpose we 
note the black sphere picture \cite{BS1} in which the black sphere radius 
$a$ is determined from $2pa\sin(\theta_M/2)=5.1356\cdots$ with the proton 
incident momentum in the center-of-mass frame, $p$, and the first peak angle 
for the measured diffraction in proton-nucleus elastic scattering, $\theta_M$. 
We then recall the known properties from the black sphere picture for 
$T_p=800$--1000 MeV \cite{BS2}: (i) the black sphere radius $a$ globally 
behaves as $1.214A^{1/3}$ fm.  (ii) the measured $\sigma_R$ agrees with 
$\pi a^2$ within error bars.  We also note that $\sigma_R\approx \pi a^2$ is 
satisfied for the values of $T_p$ down to about 100 MeV \cite{unpub} and that 
$\sigma_R$ is nearly flat at least for $T_p=100$--1000 MeV 
\cite{Bauhoff}.  It is thus natural to set
\begin{equation}
\sigma_R=\pi a_0^2 \left(1+\frac{\Delta a}{a_0}\right)^2,
\end{equation}
where $a_0$ is the black sphere radius at $T_p=800$ MeV, and $\Delta a$ is 
the deviation of $a$ from $a_0$ at given $T_p$ above 100 MeV.
In this setting, we assume that the incident protons are point particles,
leading to vanishing cross section for the proton-proton case.  This is 
reasonable since the measured proton-proton reaction cross section is
relatively small at $T_p\lesssim1000$ MeV \cite{Arndt,Hess}.

     In deriving the expression for $\Delta a$, we focus on the
``optical'' depth for incident protons.  Note that in a real situation, 
the black sphere radius $a$ corresponds to a critical radius
inside which the protons are attenuated in a target nucleus.
More explicitly, the optical depth for the protons, 
\begin{equation}
\int_l dl [\sigma_{pn}n_n({\bf r})+\sigma_{pp}n_p({\bf r})],
\end{equation}
amounts to a critical value when the nearest distance between the proton
trajectory and the nuclear target is $a$.  Here $l$ is the proton trajectory, 
$n_N({\bf r})$ is the point $N$-nucleon density, and $\sigma_{pN}$ is
the empirical $pN$ total cross section (e.g., Refs.\ 
\cite{Arndt,Hess,Ray}).  Let us now approximate the trajectory, 
although it is slightly distorted by the Coulomb repulsion, by a straight 
line.  We also approximate the nucleon distributions by a trapezoidal 
form of the length of the bottom, $R$, and that of the top, $R-D$, as shown in
Fig.\ 1, in such a way that they follow a typical behavior of the 
distributions deduced from elastic scattering data off stable nuclei 
\cite{Bat:ANP,DeV:ATO}, i.e., $n_n(r=0)+n_p(r=0)=\rho_0\equiv 0.16$ fm$^{-3}$ 
and $D=2.2$ fm.  This $D$ corresponds to the thickness parameter of about 0.53
fm in the two-parameter Fermi distribution.  We then replace the optical depth
by an effective one,
\begin{equation}
\tau={\bar \sigma}_{pN} n_c L',
\label{depth}
\end{equation}
where ${\bar \sigma}_{pN}=(Z/A)\sigma_{pp}+(1-Z/A)\sigma_{pn}$, $n_c$
is the total nucleon density at $r=a$, and $L'=2\sqrt{R^2-a^2}$ is the
length of the part of the critical trajectory in which the total nucleon
density is larger than $n_c$.  The value of $\tau$ is set to be 0.9 since 
this is consistent with the values of $a_0$ and $n_c$ for $^{12}$C, $^{58}$Ni, 
$^{124}$Sn, and $^{208}$Pb that were deduced from the elastic 
scattering data (see Ref.\ \cite{BS1} and references therein).

\begin{figure}[t]
\begin{center}
\hspace*{-0.65cm}
\includegraphics[width=10.5cm]{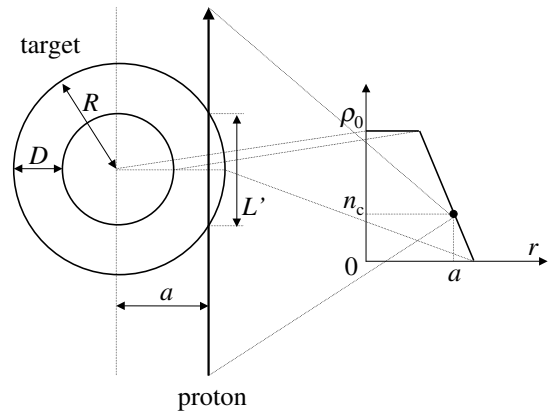}
\end{center}
\vspace*{-0.5cm}
\caption{Model for the density distribution of a target nucleus and the 
critical proton trajectory inside which the reaction with the target
nucleus occurs.} 
\end{figure}

     Let us now consider the deviation $\Delta X$ of $X={\bar\sigma}_{pN}$, 
$a$, $n_c$, and $L'$ from the value $X_0$ at $T_p=800$ MeV.  As long as 
$T_p>100$ MeV, $\Delta {\bar\sigma}_{pN}$ is sufficiently small that
\begin{equation}
   \tau \frac{\Delta{\bar\sigma}_{pN}}{{\bar\sigma}_{pN0}^2}\approx
    -n_{c0}\Delta L' -L'_0 \Delta n_c.
    \label{var1}
\end{equation}
Because of the assumed trapezoidal distribution, one can obtain
\begin{equation}
   \Delta n_c = -\frac{\rho_0}{D}\Delta a.
    \label{var2}
\end{equation}
Combining Eqs.\ (\ref{var1}) and (\ref{var2}), one can express the relation
between $\Delta{\bar\sigma}_{pN}$ and $\Delta a$ as
\begin{equation}
  \tau \frac{\Delta{\bar\sigma}_{pN}}{{\bar\sigma}_{pN0}^2}\approx
    \left(\frac{\rho_0}{D}L'_0 
        - \left.\frac{dL'}{da}\right|_0 n_{c0}\right) \Delta a.
    \label{deltaa}
\end{equation}
We thus obtain
\begin{equation}
  \sigma_R\approx\pi a_0^2 \left[1+
      \left(\frac{\rho_0 a_0}{D n_{c0}}
          -\frac{a_0}{L'_0}\left.\frac{dL'}{da}\right|_0 \right)^{-1}
  \frac{\Delta{\bar\sigma}_{pN}}{{\bar\sigma}_{pN0}}\right]^2.
  \label{final}
\end{equation}

     We now proceed to express the length $R$.  In doing so, we first
note the normalization condition for the assumed trapezoidal
distribution,
\begin{equation}
    A=\frac{4\pi \rho_0}{3}
    \left(R^3-\frac32 D R^2 + D^2 R -\frac14 D^3\right).
\end{equation}
One can seek an approximate solution to this equation by setting
$R=R_0+D/2+\delta R$, with $R_0=(3A/4\pi\rho_0)^{1/3}$, and assuming that
$\delta R$ is small.  One thus obtains
\begin{equation}     
    R\approx R_0+\frac{D}{2}-R_0\left(1+\frac{12R_0^2}{D^2}\right)^{-1}.
    \label{R}
\end{equation}
From this expression, $\delta R$ can be shown to be small even for
light elements.  Hereafter we substitute expression (\ref{R}) into
the reaction cross section formula (\ref{final}) through $L'$.  Note that 
since both $a_0$ and $R_0$ behave as $\propto A^{1/3}$, the derivative
$dL'/da$ may well be performed by assuming $R_0\propto a$.

    In Fig.\ 2 we compare the formula (\ref{final}) with the empirical data 
for the reaction cross section for targets such as C, Ca, Zr, and Pb.
The formula (\ref{final}) in which $a_0$ is set to be a value determined from 
the measured angle of the first diffraction maximum in proton elastic 
scattering \cite{BS1,BS2} well reproduces the $T_p$ dependence of the 
measured reaction cross section \cite{Bauhoff,Auce} for $T_p$ down to about 
100 MeV.  Remarkably, it agrees almost completely with the 100--180 MeV data 
updated by Auce {\it et al}.\ \cite{Auce}.  The $T_p$ dependence of such data 
is stronger for smaller $A$, a feature described by the prefactor of
$\Delta{\bar\sigma}_{pN}/{\bar\sigma}_{pN0}$ in Eq.\ (\ref{final}).  Below 100
MeV, the measured cross section shows resonance features and sharp decrease 
due to the Coulomb repulsion between the incident proton and target nucleus, 
both of which are not allowed for in constructing the present formula.  The 
neglect of the Coulomb corrections is reasonable above 100 MeV, since the 
empirical reaction cross section for protons on stable nuclei is appreciably 
lower than that for neutrons only at incident energy lower than $\sim70$ MeV 
\cite{Chadwick}.  Below 100 MeV, there is another limitation that faces 
the formula (\ref{final}): $\sigma_{pn}$ is too large to make the formula 
applicable \cite{Hess}.  We remark that the data above 200 MeV, tabulated 
in Ref.\ \cite{Bauhoff}, have to be seen with caution.  This is partly because
the recent data at energies between 80 and 180 MeV \cite{Auce} are not 
smoothly connected with the data above 200 MeV and partly because the data 
between 200 MeV and 600 MeV are systematically larger than the data at 860 
MeV, in contrast to the tendency of the nucleon-nucleon total cross section 
\cite{Arndt,Hess}.  We also remark that the previous predictions based on 
microscopic optical potential models \cite{Ernst,Ray,Dirac} are more sensitive
to $T_p$ than ours in a way more difficult to reproduce the $T_p$ dependence 
of the data.  In fact, such models involve the exponential $\sigma_{pN}$
dependence in the prediction of the reaction cross section through the phase 
shift function, in contrast to the power-law  $\sigma_{pN}$ dependence of
the formula (\ref{final}).  The importance of the power-law $\sigma_{pN}$ 
dependence is empirically known \cite{Ing}.

\begin{figure}[t]
\begin{center}
\hspace*{-0.2cm}
\includegraphics[width=4.37cm]{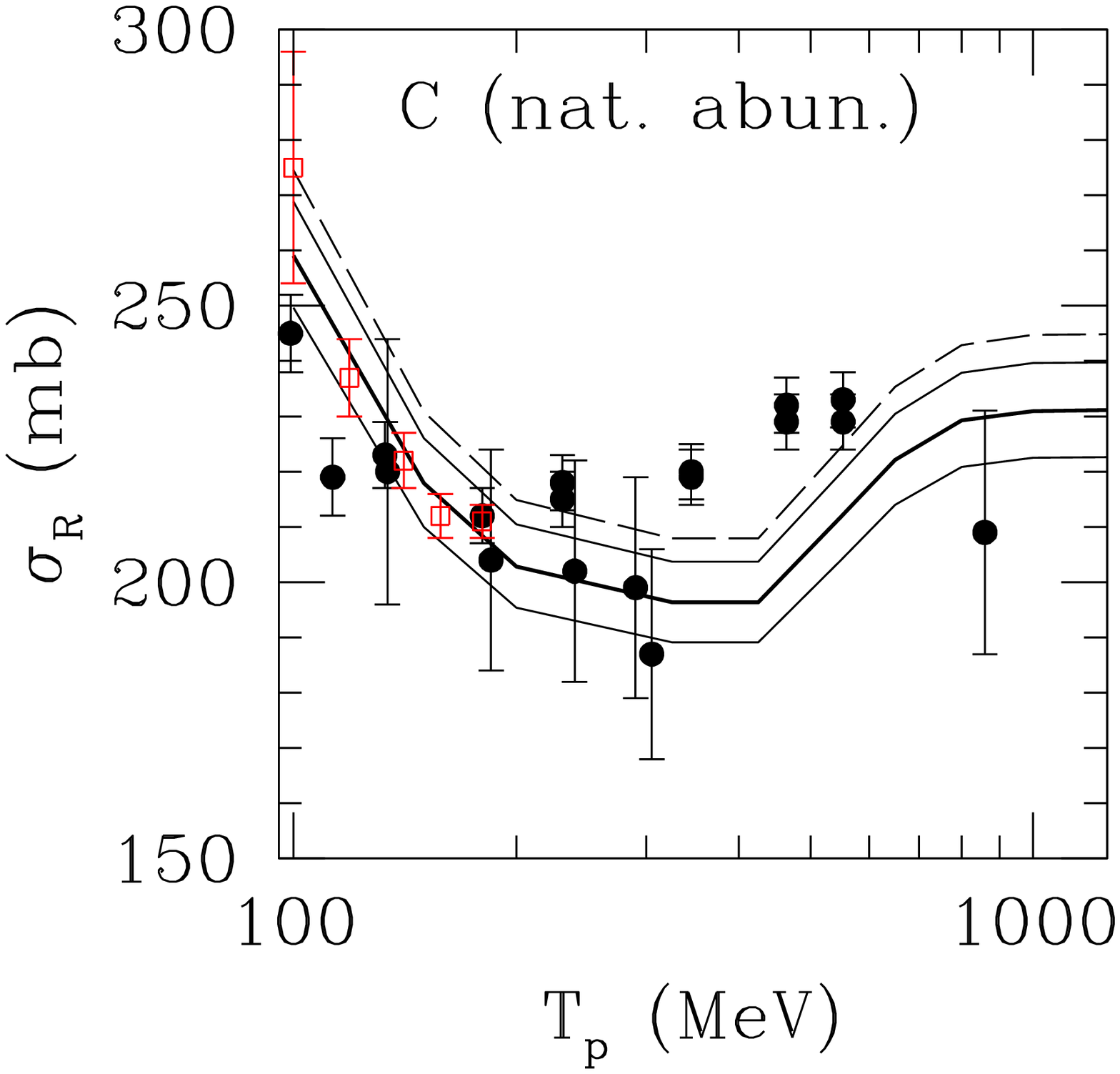}
\hspace*{-0.2cm}
\includegraphics[width=4.37cm]{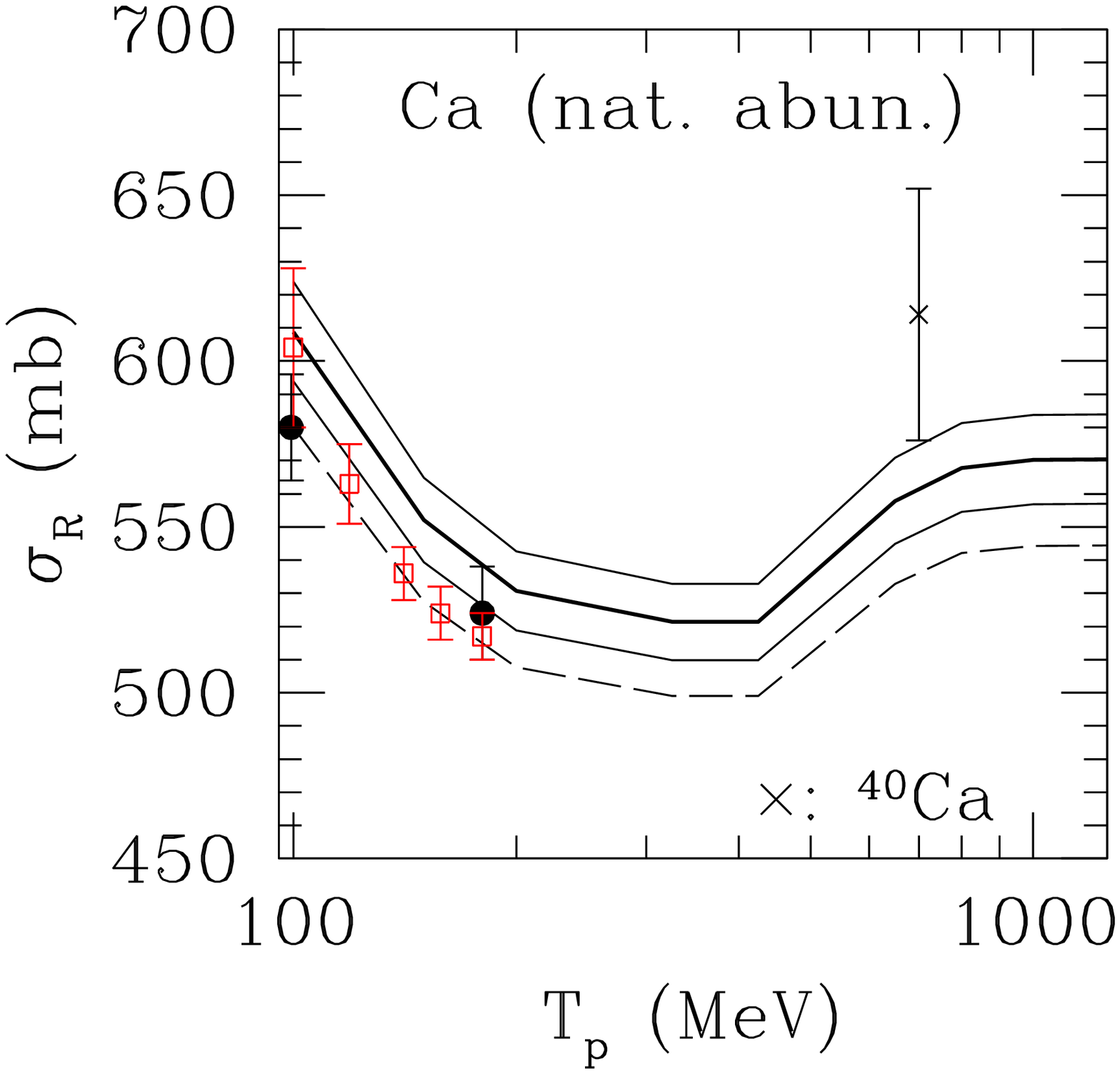}
\hspace*{-0.2cm}
\includegraphics[width=4.37cm]{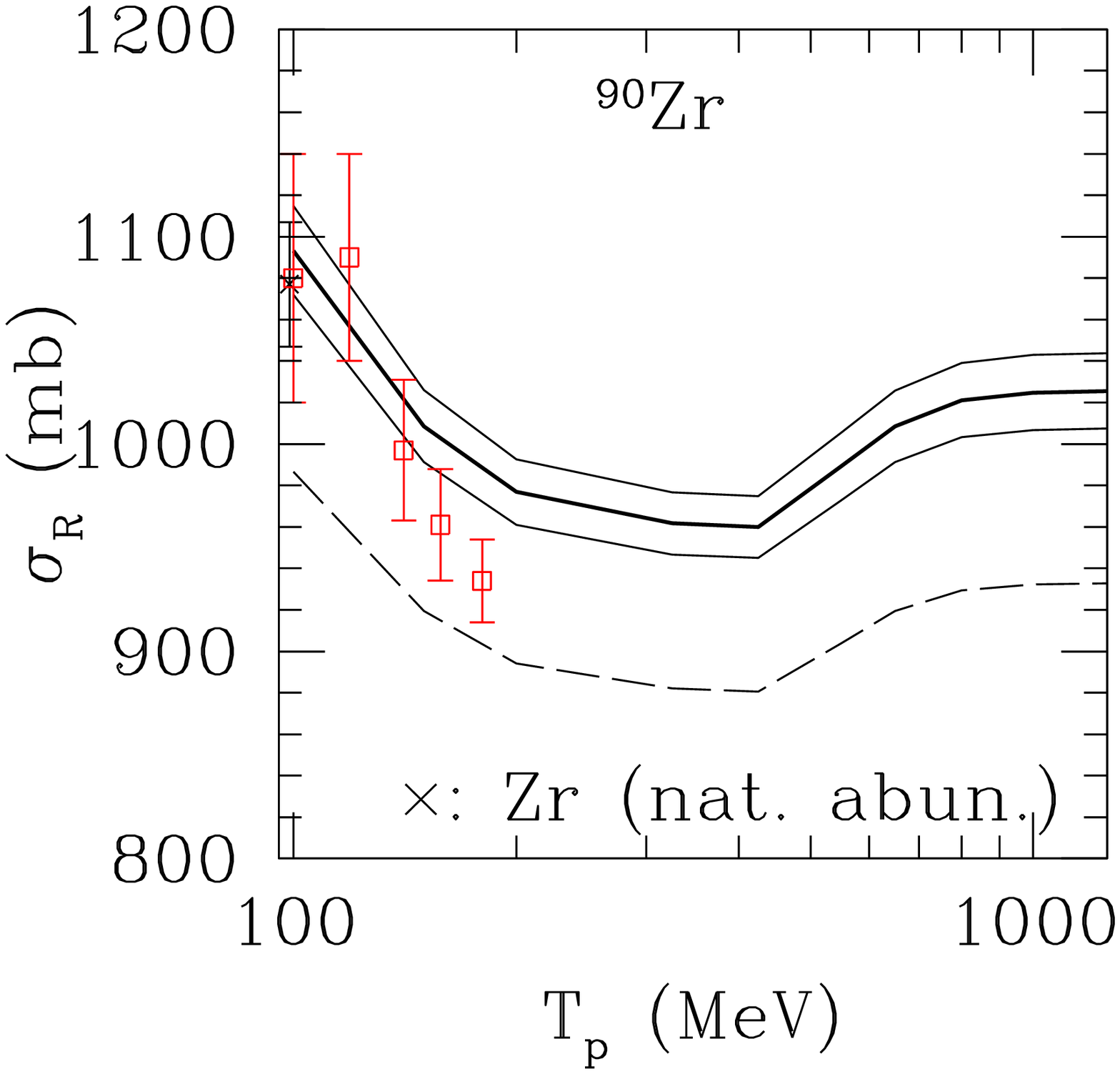}
\hspace*{-0.2cm}
\includegraphics[width=4.37cm]{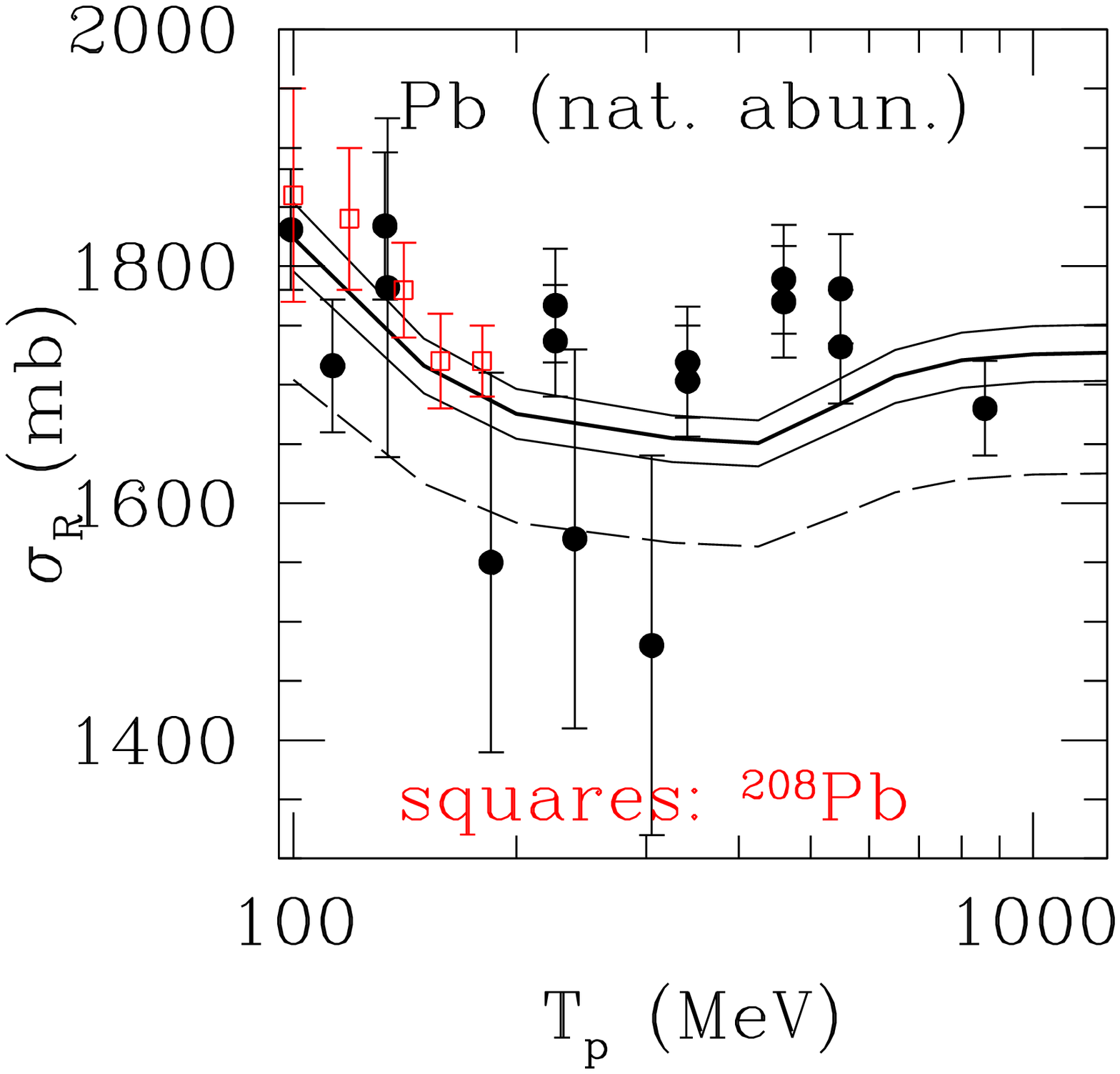}
\end{center}
\vspace{-0.5cm}
\caption{(Color online) The proton-nucleus reaction cross section as a 
function of proton incident energy.  The target nucleus is natural C, 
natural Ca, $^{90}$Zr, and natural Pb.  The squares denote 
the data from Auce $et$ $al.$ \cite{Auce}, while the filled circles and 
crosses denote the data tabulated in Ref.\ \cite{Bauhoff}.  Note some 
exceptions in which the target nucleus is slightly different in mass number: 
crosses for $^{40}$Ca and natural Zr and squares for $^{208}$Pb.
The lines are the results from the formula (\ref{final}) in which $a_0$ is 
set to be 1.214$A^{1/3}$ fm (dashed lines),
2.7 fm (bold line for C), 2.65 fm (lower, thin line for C), 2.75 fm (upper, 
thin line for C), 4.25 fm (bold line for Ca), 4.2 fm (lower, thin line for Ca),
4.3 fm (upper, thin line for Ca), 5.7 fm (bold line for Zr), 5.65 fm (lower, 
thin line for Zr), 5.75 fm (upper, thin line for Zr), 7.4 fm (bold line for
Pb), 7.35 fm (lower, thin line for Pb), and 7.45 fm (upper, 
thin line for Pb).  The choice of 2.65--2.75 fm for C, 4.2--4.3 fm for Ca,
5.65--5.75 fm for Zr, and  7.35--7.45 fm for Pb is based on the measured 
diffraction peak angle in proton elastic scattering.
} 
\end{figure}

    From Fig.\ 2 we can also examine to what extent a simple choice of $a_0$ 
as $1.214A^{1/3}$ fm works in the cross section formula.  As was seen from 
Ref.\ \cite{BS2}, this simple scaling well reproduces the values of $a_0$
determined from the measured angle of the first diffraction maximum in 
proton elastic scattering for stable nuclei ranging from light to heavy 
elements.  In fact, we see a good agreement between the results for the 
cross section formula from both choices of $a_0$, although the values 
from the simple $A^{1/3}$ scaling tend to be lower than those from the other
by typically 5--10 \% for $A\gtrsim100$.

     As far as the $A$ dependence is concerned, the present formula is similar
to the parametrization constructed by Kox $et$ $al.$ \cite{Kox}.  However, 
they differ significantly in the way of describing the energy dependence.  Kox
$et$ $al.$ introduced a semiempirical energy dependent parameter that takes 
care of changing ``surface transparency'' as the projectile energy 
changes, while the energy dependence of our formula stems solely from that of 
${\bar \sigma}_{pN}$ and hence our formula is free from any adjustable energy 
dependent parameter.  We remark that the formula by Kox $et$ $al.$ is 
applicable for a wider range of the incident energy including the low energy 
region where the Coulomb repulsion between the projectile and target takes
effect, although it ignores resonance features which are also important
in such a low energy region.  We also remark that there is such a kind of
formula as includes two adjustable energy dependent parameters 
\cite{Sil,Carl,Mach}.

     We now extend the cross section formula (\ref{final}) constructed for
stable nuclei to unstable nuclei by introducing the dependence on the neutron 
excess $\delta$.  In doing so, we first note a formula for the rms radii
of proton and neutron point distributions, $R_p$ and $R_n$, which reads
\begin{equation}
R_p=c_1 A^{1/3}+c_2+c_3(\delta-\delta_0)^2,
\label{rpfit}
\end{equation}
with $c_1=0.915$ fm, $c_2=-0.102$ fm, $c_3=0.389$ fm, and $\delta_0=0.880$, and
\begin{equation}
R_n=c_4 A^{1/3}(1+c_5 L\delta^2+c_6 L^2\delta^4)+c_7+c_8\delta,
\label{rnfit}
\end{equation}
with $c_4=0.880$ fm, $c_5=0.00635$ MeV$^{-1}$, $c_6=-0.000172$ MeV$^{-2}$, 
$c_7=0.302$ fm, and $c_8=0.193$ fm.  This formula was constructed in Ref.\ 
\cite{OI} by fitting the calculations obtained from macroscopic nuclear models
for $A>50$, $0<\delta<0.3$, and various values of the density symmetry 
coefficient $L$ and the incompressibility $K_0$ of symmetric nuclear matter.
It was noted that the calculated results for the rms radii are almost 
independent of $K_0$.  We remark that the formulas (\ref{rpfit}) and 
(\ref{rnfit}), when extrapolated into the range of $4\leq A \leq 50$, 
reproduces as well the values of the rms charge and matter radii deduced from 
elastic scattering data for stable nuclei.

     The formulas (\ref{rpfit}) and (\ref{rnfit}) imply the significance
of allowing for uncertainties in the density symmetry coefficient $L$ in 
describing the size dependent quantities at finite neutron excess.  In fact, 
$L$ acts to increase the nuclear size at nonzero $\delta$.  This is mainly 
because the saturation density of uniform nuclear matter at finite neutron 
excess decreases from that in the symmetric case in such a way that the 
decrease is roughly proportional to $L\delta^2$ \cite{OI}.  This effect of 
$L$ could be seen by systematically analyzing the first diffraction maxima
in the proton elastic scattering data \cite{IOB} and thus could manifest 
itself more remarkably in the reaction cross section which behaves as the size
squared.

     In order to describe the cross section at finite neutron excess by
including the effect of $L$, we multiply expression (\ref{final}) by a scale 
factor $f(A,\delta; L)=R_m^2(A,\delta; L)/R_m^2(A,\delta=0; L)$ 
with the rms matter radius $R_m=\sqrt{(N/A)R_n^2+(Z/A)R_p^2}$.  This is 
reasonable, because, at least for stable nuclei heavier than $A\sim50$, 
the rms matter radii deduced in previous elaborate analyses from elastic 
scattering data agree very well with $\sqrt{3/5}a$ \cite{BS1}.  Although this 
agreement neither holds for $A\lesssim 50$ \cite{BS2} nor is obvious at large 
neutron excess, it is natural to expect that the isospin dependence of $R_m$ 
is similar to that of $a$.  We remark in passing that the scale factor $f$ 
is sufficiently close to unity for stable nuclei including $^{208}$Pb that 
it does almost no damage to the good agreement between the formula and the 
data.

      In addition to the isospin dependence coming through this scale factor, 
another isospin dependence is inherent in the formula (\ref{final}) through 
the part associated with ${\bar\sigma}_{pN}$.  In the absence of detailed 
information about the surface diffuseness and neutron skin at large neutron 
excess, we simply leave it as it is.  This is expected to be good enough since
$\sigma_{pp}$ and $\sigma_{pn}$ are not so different
at $T_p=100$--1000 MeV as to affect the analysis of the reaction cross section 
data within the error bars, which are typically of order a few percent.

      In the black sphere picture adopted here, possible neutron halos have 
no influence on the construction of the cross section formula for unstable
nuclei.  This is because such halos correspond to a very low density region 
which is ``optically'' very thin.  Moreover, the macroscopic nuclear 
models leading to Eqs.\ (\ref{rpfit}) and (\ref{rnfit}) do not allow for the 
possible presence of a neutron halo \cite{OI}.  If the cross section formula 
is significantly smaller than the empirical value for halo nuclei for 
reasonable values of $L$, therefore, this would signify the limitation of  
the present framework and the enhancement of the reaction cross section due 
to the presence of halos.

      We now focus on the interaction cross section data for a halo nucleus
$^{11}$Li \cite{Tani92} as well as the reaction cross section data for helium 
isotopes \cite{Neumaier1}.  Note that the interaction cross section can 
be regarded as equal to the reaction cross section for halo nuclei in which 
internucleon excitations generally involve breakup of halo neutrons. 
Such light elements are expected to be suitable for examining the 
isospin dependence, partly because for stable nuclei $a_0$ is consistent with 
$1.214 A^{1/3}$ fm, and partly because $f$ can amount to about 1.1.
In fact, the formula (\ref{final}) with $a_0=1.214 A^{1/3}$ fm well
reproduces the data for $^{4}$He \cite{Jaros} and Li of the natural abundance
\cite{Jo61}.  From the formula (\ref{final}) further multiplied by the 
scale factor $f$ with a typical $L$ value of 100 MeV, we obtain a smaller 
value for $^{11}$Li by about 10 \%, but a value consistent with the empirical 
data for halo nuclei $^{6}$He and $^{8}$He.  It might be tempting to 
attribute this deviation for $^{11}$Li to the presence of a neutron halo, 
since the formula (\ref{final}) and the scale factor $f$ are not influenced
by an extremely low density tail of the neutron distribution.  However, this
attribution is not obvious because of the consistency obtained for $^{6}$He 
and $^{8}$He which are known to have a similar three-body structure to that of
$^{11}$Li \cite{Chulkov}.

     So far we have assumed that nuclei are spherical.  Nuclear deformation,
however, does affect the reaction cross section and thus might be 
deduced from the experimental data measured for heavy deformed nuclei.
For simplicity, we focus on quadrupolar deformation and describe a
nucleus as a static ellipsoid having the length, $b_1$, $b_1$, $b_2$, of the 
three axes.  Both for prolate ($b_1<b_2$) and oblate ($b_1>b_2$) deformations
with random orientations, the cross section seen by incident protons becomes,
on the average, larger than the value for a sphere of the same volume.  
The relative increment can be calculated as
\begin{equation}
  g=\frac25 \alpha_2^2 + \cdots,
\end{equation}
where $\alpha_2$ is the quadrupolar deformation factor satisfying
$b_2/b_1=(1+\alpha_2)/(1-\alpha_2/2)$.  This suggests that even for 
large deformation $|\alpha_2|\sim0.2$, $g$ amounts to about 2 \% only.

     In the above estimate of the effect of deformation, we ignore possible 
rotation of the nucleus.  Since the rotation speed is far smaller than
the speed of light, it is reasonable to consider that the incident protons, 
which move relativistically, see the target nucleus nonrotating during the 
flight.  However, rotation would tend to enlarge the nucleus itself.  Balance 
between the centrifugal force and the nuclear incompressibility would 
typically give rise to about a percent enhancement of the cross section.
We can thus conclude that the combined effect of deformation and rotation
is significantly smaller than the effect of large neutron excess.

\section{Formula for nucleus-nucleus reaction cross section}

     In this section we generalize the formula constructed in Sec.\ 
\ref{fpA} to the case of nucleus-nucleus reaction cross section.  As in 
Ref.\ \cite{BS2}, we straightforwardly set it as
\begin{equation}
   \sigma_R=\pi(a_P+a_T)^2,
\end{equation}
where $a_P$ and $a_T$ are the black sphere radii of the projectile 
($Z_P$, $A_P$) and target ($Z_T$, $A_T$) nucleus.  By following a line of 
argument for the proton-nucleus case, we obtain
\begin{equation}
    \sigma_R=\pi\left[(a_{P0}+\Delta a_P)f_P^{1/2}
                     +(a_{T0}+\Delta a_T)f_T^{1/2}\right]^2,
    \label{AA}
\end{equation}
where $f_P$ and $f_T$ are the scale factors for the projectile and target 
nucleus, and $\Delta a_P$ and $\Delta a_T$ are the deviations of $a_P$ and 
$a_T$ from the values of $a_{P0}$ and $a_{T0}$ at incident energy
per nucleon of 800 MeV, both of which are determined from Eq.\ 
(\ref{deltaa}) in which ${\bar \sigma}_{pN}$ is replaced by
\begin{eqnarray}
{\bar \sigma}_{NN}&=&\left(\frac{Z_T}{A_T}\right)
\left[\left(\frac{Z_P}{A_P}\right)\sigma_{pp}+\left(1-\frac{Z_P}{A_P}\right)
\sigma_{pn}\right]
\nonumber \\ & &
 +\left(1-\frac{Z_T}{A_T}\right)\left[\left(\frac{Z_P}{A_P}\right)\sigma_{pn}
 \right.\nonumber \\ & & \left.
 +\left(1-\frac{Z_P}{A_P}\right)\sigma_{pp}\right]
\end{eqnarray}
for $\Delta a_T$ and ($T \leftrightarrow P$) for $\Delta a_P$.  Here we have 
omitted the effects of nuclear deformation.  The formula (\ref{AA}) reduces to
the proton-nucleus case by setting either $a_P$ or $a_T$ to be zero.  
Hereafter we will set $a_{P0}=1.214A_{P}^{1/3}$ fm and 
$a_{T0}=1.214A_{T}^{1/3}$ fm unless notified otherwise.

     The formula (\ref{AA}) works well for the $^{12}$C-$^{12}$C case for 
energy per nucleon at least down to 200 MeV (see Fig.\ 3).  In fact, in this 
energy range the empirical values of the $^{12}$C-$^{12}$C case \cite{Kox}
are almost four times as large as those of the $p$-$^{12}$C case, while the 
formula (\ref{AA}) for the $^{12}$C-$^{12}$C case is just four times as large 
as that for the $p$-$^{12}$C case.  At lower energies, the formula (\ref{AA}) 
overestimates the reaction cross section to an extent much larger than the 
one that could be explained by our neglect of the Coulomb repulsion.  
This suggests the necessity of modifying the classical optical depth
argument underlying the formula (\ref{AA}) in analyzing the low energy 
nucleus-nucleus reactions.  The agreement for energy per nucleon down to 200 
MeV and the deviation for lower energy can also be seen in all the other 
reactions between stable nuclei tabulated in Table II of Ref.\ \cite{Kox}.

\begin{figure}[t]
\begin{center}
\includegraphics[width=8.4cm]{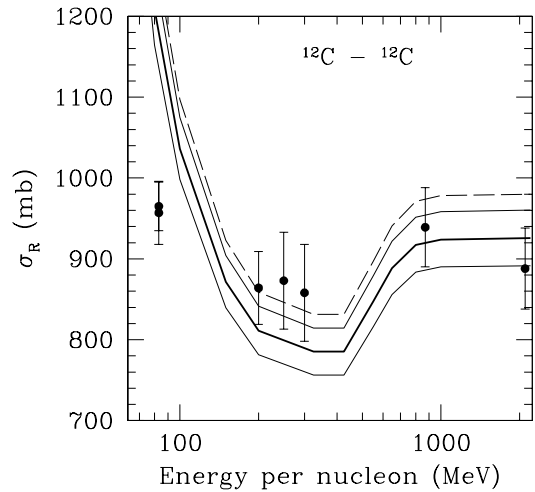}
\end{center}
\vspace{-0.5cm}
\caption{The $^{12}$C-$^{12}$C reaction cross section as a function of energy 
per nucleon.  The filled circles are the empirical data from Refs.\ 
\cite{Kox,Jaros,Zheng}.  The lines are the results from the 
formula (\ref{AA}) in which $a_0$ is set to be 2.7 fm (bold), 2.65 fm 
(lower, thin), 2.75 fm (upper, thin), and 1.214$A^{1/3}$ fm (dashed).
The choice of 2.65--2.75 fm is based on the measured diffraction peak angle 
in proton elastic scattering.
} 
\end{figure}

\begin{figure}[t]
\begin{center}
\includegraphics[width=8.4cm]{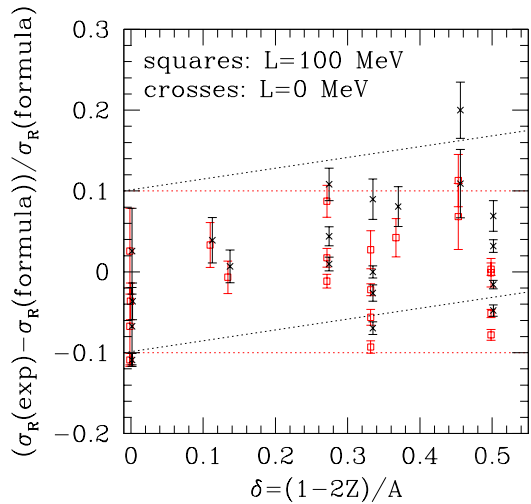}
\end{center}
\vspace{-0.5cm}
\caption{(Color online) Deviation of the empirical reaction cross section from
the cross section formula (\ref{AA}) with $L=0$ MeV (crosses) and 
$L=100$ MeV (squares).  $a_{P0}$ and $a_{T0}$ are set to be
$1.214A_{P}^{1/3}$ fm and $1.214A_{T}^{1/3}$ fm, respectively.
The dotted lines are drawn to guide the eye.} 
\end{figure}

     We note that for the $^4$He-$^{12}$C case, the formula (\ref{AA})
significantly overestimates the reaction cross section even at energy per 
nucleon of order or higher than 800 MeV as can be seen by comparison 
with the empirical data \cite{Jaros}.  This exceptional behavior is 
attributable to the fact that excitations associated with internucleon motion 
are highly suppressed in $\alpha$ particles.

     It is important to compare the constructed formula with the empirical
data for unstable nuclei incident on a stable nucleus target.  
We note the data for the interaction cross section at energy per nucleon 
of order 800 MeV tabulated in Ref.\ \cite{Ozawa}.  For candidates of halo 
nuclei such as $^6$He, $^8$He, $^{11}$Li, $^{11}$Be, and $^{19}$C, we may 
assume that the interaction cross section is almost equal to the total 
reaction cross section.  For a typical value of $L$ of order 100 MeV,
we obtain a good agreement between the formula and the data for Be, C, and 
Al targets; the difference is of order 5 \%, 
typically.  The results from these comparisons, together
with the aforementioned proton-nucleus cases ($^{11}$Li \cite{Tani92},
$^{4}$He \cite{Neumaier1,Jaros}, $^6$He \cite{Neumaier1}, $^8$He 
\cite{Neumaier1}, natural Li \cite{Jo61}, $^9$Be \cite{Jo61}, natural C 
\cite{Jo61}), are summarized in Fig.\ 4.  We find that the deviation
between the empirical data and the formula with $L=100$ MeV
is within $\pm10$ \%, while the formula with $L=0$ MeV tends to underestimate
the cross section at large neutron excess.  Judging from the scattering 
pattern in the plot for $L=100$ MeV, which is almost uniform within the 
band between $\pm10$ \%, and recalling that the formula is not influenced
by the presence of neutron halos, we see no absolute need for taking into 
account the possible enhancement of the reaction cross section by an extremely
low density tail of the neutron distribution in explaining the empirical 
values of the reaction cross section.  In the absence of this 
enhancement, it is expected that one can deduce the value of $L$ from the 
comparison between the empirical data and the formula.  We note, however, 
that the difference of the formula with $L=100$ MeV from the formula with 
$L=0$ MeV is of order or even smaller than that from the empirical data.  
This suggests the necessity of a systematic analysis in deducing the value 
of $L$ itself.

\section{Conclusion}

     We constructed a formula for the proton-nucleus reaction cross section 
in a way free from any adjustable $T_p$ dependent parameter.  The
dependence of the cross section on $A$ and $T_p$ was deduced by combining
the black sphere picture of nuclei with a simple argument of the proton
optical depth.  For stable nuclei, this formula remarkably well reproduces 
the empirical $T_p$ dependence of the reaction cross section at 
$T_p=100$--1000 MeV.  This suggests a great advantage over previous 
microscopic optical potential models in which the $T_p$ dependence becomes too
strong through the phase shift function to reproduce the empirical 
behavior.  In addition, we took into account the effects of large neutron 
excess typical of unstable nuclei and nuclear deformation on the reaction
cross section.  We thus found out a possible way of examining the 
influence of the density dependence of the symmetry energy and the presence 
of neutron halos on the reaction cross section.

\section*{Acknowledgments}

     We acknowledge the members of Japan Charged-Particle Nuclear Reaction 
Data Group (JCPRG), especially N. Ohtsuka, for kindly helping us collect 
various data sets.

\end{document}